\documentclass[aps,twocolumn,groupedaddress,superscriptaddress,amsmath,amssymb,prl,longbibliography]{revtex4-2}
\usepackage{mathtools,amsmath,amsxtra}
\usepackage{amssymb}
\usepackage{physics}
\usepackage[english]{babel}
\usepackage{dsfont}
\usepackage{graphicx}
\usepackage{dcolumn}
\usepackage{bm}
\usepackage{bbold}
\usepackage[usenames,dvipsnames]{color}
\usepackage[colorlinks,citecolor=Blue,linkcolor=Red, urlcolor=Blue]{hyperref}
\usepackage{tikz}
\usetikzlibrary{decorations.pathmorphing}
\usetikzlibrary{arrows.meta}
\usepackage{braket}
\usepackage[normalem]{ulem}
\usepackage{lipsum}
\usepackage{caption}
\DeclareCaptionType{equ}[][]

\newcommand{\ste}[1]{\textcolor{green}{#1}}

\begin{document}

\title{Fault-Tolerant Computing with Single Qudit Encoding}

\author{M. Mezzadri$^\dagger$}
\affiliation{Università di Parma, Dipartimento di Scienze Matematiche, Fisiche e Informatiche, I-43124, Parma, Italy}
\affiliation{Gruppo Collegato di Parma, INFN-Sezione Milano-Bicocca, I-43124 Parma, Italy}

\author{A. Chiesa$^\dagger$}
\affiliation{Università di Parma, Dipartimento di Scienze Matematiche, Fisiche e Informatiche, I-43124, Parma, Italy}
\affiliation{Gruppo Collegato di Parma, INFN-Sezione Milano-Bicocca, I-43124 Parma, Italy}
\affiliation{UdR Parma, INSTM, I-43124 Parma, Italy}

\author{L. Lepori}
\affiliation{Università di Parma, Dipartimento di Scienze Matematiche, Fisiche e Informatiche, I-43124, Parma, Italy}
\affiliation{Gruppo Collegato di Parma, INFN-Sezione Milano-Bicocca, I-43124 Parma, Italy}

\author{S. Carretta}
\email{stefano.carretta@unipr.it}
\affiliation{Università di Parma, Dipartimento di Scienze Matematiche, Fisiche e Informatiche, I-43124, Parma, Italy}
\affiliation{Gruppo Collegato di Parma, INFN-Sezione Milano-Bicocca, I-43124 Parma, Italy}
\affiliation{UdR Parma, INSTM, I-43124 Parma, Italy}

\begin{abstract}
We discuss stabilizer quantum-error correction codes implemented in a single multi-level qudit to avoid resource escalation typical of multi-qubit codes. These codes can be customized to the specific physical errors on the qudit, effectively suppressing them. We demonstrate a Fault-Tolerant implementation on molecular spin qudits, showcasing nearly exponential error suppression with only linear qudit size growth. Notably, this outperforms qubit codes using thousands of units. We also outline the required properties for a generic physical system to Fault-Tolerantly implement these embedded codes.
\end{abstract}

\maketitle

\twocolumngrid 
A ‘‘useful" quantum computer should be universal, accurate and scalable \cite{Chao2018}.
Meeting these conditions on a noisy hardware requires a Fault-Tolerant quantum error correction (QEC) approach, which allows for universal quantum computation with an arbitrary error reduction, while remaining experimentally achievable.\\
The standard way to tackle this problem relies on encoding the elementary unit of computation (a \textit{logical qubit}, LQ) into a collection of distinct physical qubits \cite{Gottesman2000,Devitt2013,Nigg2014,Terhal2015,Campbell2017}.
Different codes were developed along these lines, progressively increasing the maximum tolerated error on each elementary operation
(the so-called {\it threshold}) \cite{Aliferis2005,Knill2005,Cross2009,Wang2011,Fowler2012,Mariantoni2012,Stephens2014,Chamberland2016,Marks2017,Campbell2014}. However, all of them show an explosion in the number of physical qubits and gates to achieve the error suppression required for a reliable computation \cite{Mariantoni2012,Suchara2013,Paetznick2023}, which represents an important road-block to an actual implementation.

We pursue an alternative approach, in which the elementary protected unit of information is encoded in a single $d-$levels qudit. 
This idea of {\it embedded codes} \cite{Pirandola2008,Cafaro2012,PRXGirvin,JPCLqec,Cai2021,Chiesa2022}
would dramatically reduce the resource overhead in their physical implementation and hence has recently inspired proposals based on systems with an intrinsic multi-level structure, such as 
photons \cite{PRXGirvin,Cai2021,omanakuttan2023} or molecular spins \cite{JPCLqec,npjQI,Carretta2021,ChizziniPCCP,Lim2023}. In some cases the break-even point  has been reached \cite{Ni2023} and gates \cite{Ma2020,Reinhold2020,Puri2020} or error detection \cite{Rosenblum2018} without error propagation \cite{Hu2019,Cai2021} were proposed.
Yet, a comprehensive Fault-Tolerant implementation accommodating realistic noise models, which encompass noisy ancillae, encodings, measurements and especially noise during gates, is still missing \cite{Grimsmo2020, Cai2021}.

Here we develop the first fault-tolerant (FT) implementation of an embedded stabilizer code, in which all quantum computing (QC) steps  (i.e. a universal set of logical operations, stabilization and correction) are {\it transparent} to the errors handled by the code, i.e. errors do not propagate during any of these operations.

We demonstrate FTQC on a physical hardware consisting of Molecular Spin Qudits (MSQs), based on a physically motivated error model where pure dephasing is largely dominant \cite{JPCLqec,npjQI,Carretta2021}. We first introduce a illustrative MSQ hardware and its coupling with the environment \cite{Chiesa2022}. Thanks to a clear hierarchy in the error operators, a code subspace protected from the most relevant ones can be derived.
Then, we prove that all the QC procedures (one- and two-qubit logical gates, stabilization and recovery) can be made Error Transparent (ET) by proper pulse sequences.

We provide a threshold analysis based on  numerical simulations, showing that the corrected qubit outperforms the uncorrected one for reasonable duration of the elementary gates and of the error rate.
{\it Moreover, we find an almost exponential error reduction which comes only with a linear growth in the number of qudit levels.}

The basic ingredients which grant FTQC 
are (i) a well defined hierarchy in the errors affecting the physical hardware and (ii) the high connectivity between its  eigenstates. 
This latter feature can be rather easily obtained in MSs, thanks to their unparalleled tunability at the synthetic level \cite{Sessoli2019,Carretta2021}. 
Nevertheless, the code can be extended to other qudit-based hardwares meeting the requirements we have identified for FTQC.

This approach extremely simplifies the actual control of the hardware and could make the \textit{qudit-embedded} approach a viable path towards the physical implementation of a ‘‘useful" quantum computing machine.

{\it General protocol --}
Errors corrupting quantum information initially encoded in the pure state $\rho_0 = \ket{\psi_0} \bra{\psi_0}$ are represented by Kraus operators $E_k$, i.e. the density matrix after possible errors is $\rho = \sum_k E_k \rho_0 E_k^\dagger$.
Protection against a set of errors $\{ E_k \}$ can be achieved by identifying code words $\ket{0_L}$ and $\ket{1_L}$ which satisfy Knill-Laflamme conditions (KLc) \cite{KnillLaflamme}:
\begin{subequations}
\begin{eqnarray} 
    \bra{0_L}{E}_k^\dagger E_j\ket{0_L} &=& \bra{1_L}{E}_k^\dagger E_j\ket{1_L} \\
    \bra{0_L} {E}_k^\dagger E_j\ket{1_L} &=& 0 .
\end{eqnarray}
\label{eq:KL}
\end{subequations}
By using $d$ dimensional qudits, at most $d/2$ different errors can be corrected. 
We then consider the vector set $\{E_k\ket{0_L}, E_k\ket{1_L}\}_{k=0}^{d/2 -1}$, and the orthonormal basis set $\mathcal{A}$ obtained from its Gram-Schmidt orthogonalization.
We denote the set $\mathcal{A}$ as $\{\ket{\ell,k}\}$ with $\ell = 0,1$ and $k = 0, \ldots, d/2 -1$. 

Formally, given a gate $G$ on a single qubit, the corresponding logical gate $G_L$ that extends $G$ to the protected \textit{logical qubit} is given by $G_L \doteq G \otimes \mathrm{I}_{d/2}$.
This means applying the gate $G$ to each subspace $\ket{\ell,k}$ independently of $k$ and  highlights the basic idea to achieve FTQC, i.e. to perform logical operations \emph{independently} of the error \cite{Ma2020}.
Two-qubit controlled logical gates $CG$ can be realized in a similar way, by implementing $CG$ independently of the error $k$ on both the control and target LQs (see \ste{SM}).

Error detection translates into the stabilization of information through the measurement of  
\begin{equation}
\label{eq:stabilizer}
S = \sum_{k,\ell} \lambda_k\ket{\ell,k}\bra{\ell,k}
\end{equation}
where $\lambda_k \ne \lambda_{k'}$ for $k\ne k'$. 
This is an extension of stabilizers codes on qubits, with the important simplification that we have only one multi-valued stabilizer that gives directly the error syndrome instead of multiple two-valued stabilizers \cite{Fowler2012b,Bravyi2014}.
This reduces the impact of measurement errors (see below).
Even more importantly, this \textit{completely removes the need for a syndrome decoder}, which is usually an important bottleneck for large qubit stabilizer codes \cite{Das2022,Sweke2021}. 

Stabilization is achieved by exploiting a $d/2-$levels ancilla linked to the $d-$levels LQ and implementing a $k$-controlled operation $CU$ between the LQ (control, initially in a generic state $\ket{\bar{\psi}}$) and the ancilla (target, initialized in its ground state $\ket{0}$). 
The $CU$ gate acts as follows: $\ket{\ell,k}\ket{0} \mapsto \ket{\ell,k}\ket{k}$, $\ket{\ell,k}\ket{k} \mapsto -\ket{\ell,k}\ket{0}$ and the identity otherwise 
\footnote{The relative phase between $\ket{\ell,k}\ket{0}$ and $\ket{\ell,k}\ket{k}$ is necessary to ensure that the corresponding generator has zero diagonal. Since the ancilla is initialized in $\ket{0}$, this relative phase is irrelevant.}.
Therefore, $CU$ maps each error $k$ to a specific eigenstate of the ancilla. Hence, a final measurement of the ancilla in its eigenbasis identifies the error syndrome $k$ and {\it stabilizes information in the subspace $\ket{\ell,k}$}.
Note that $CU$ is ET  since it does not affect the state of the LQ but only the state of the ancilla.

The recovery (correction) step $R^L_{k}$ simply consists in mapping back the stabilized states $\ket{\ell,k}$ into $\ket{\ell,0}$ by a unitary transformation depending on $k$. Since $R^L_{k}$ can be written as 
$\mathrm{I}_2\otimes\mathrm{R_k}$, 
no evolution between subspaces with different $\ell$ is performed.

Finally, readout of the LQ state corresponds to a measurement of $Z_L$, which again translates into distinguishing the value of $\ell$ independently of the error $k$.
To simplify this step one can choose $\ell=0$ and $\ell=1$ logical subspaces disjoint in the eigenbasis.
In this way, the readout simply corresponds to summing the probabilities of finding the LQ into the eigenstates belonging to each of the two subspaces.

At last, LQ encoding can be achieved analogously to standard qubit stabilizer codes \cite{Cross2009,Aliferis2005}.
Starting from the qudit into its ground eigenstate, we drive it to the proper superposition of eigenstates defining  $\ket{0_L}\equiv \ket{0,0}$. 
Errors occurring during this step will bring the system to a mixture $\sum_{\ell \ell' k k'} \rho_{\ell,k}^{\ell',k'} \ket{\ell,k} \bra{\ell', k'}$.
Hence, we first perform a logical measurement of the qudit discarding it if we get an $\ell = 1$ outcome and we finally proceed with a stabilization step discarding it for any $k\ne0$.

\begin{figure*}[ht!]
    \centering
    \includegraphics[width=0.95\textwidth]{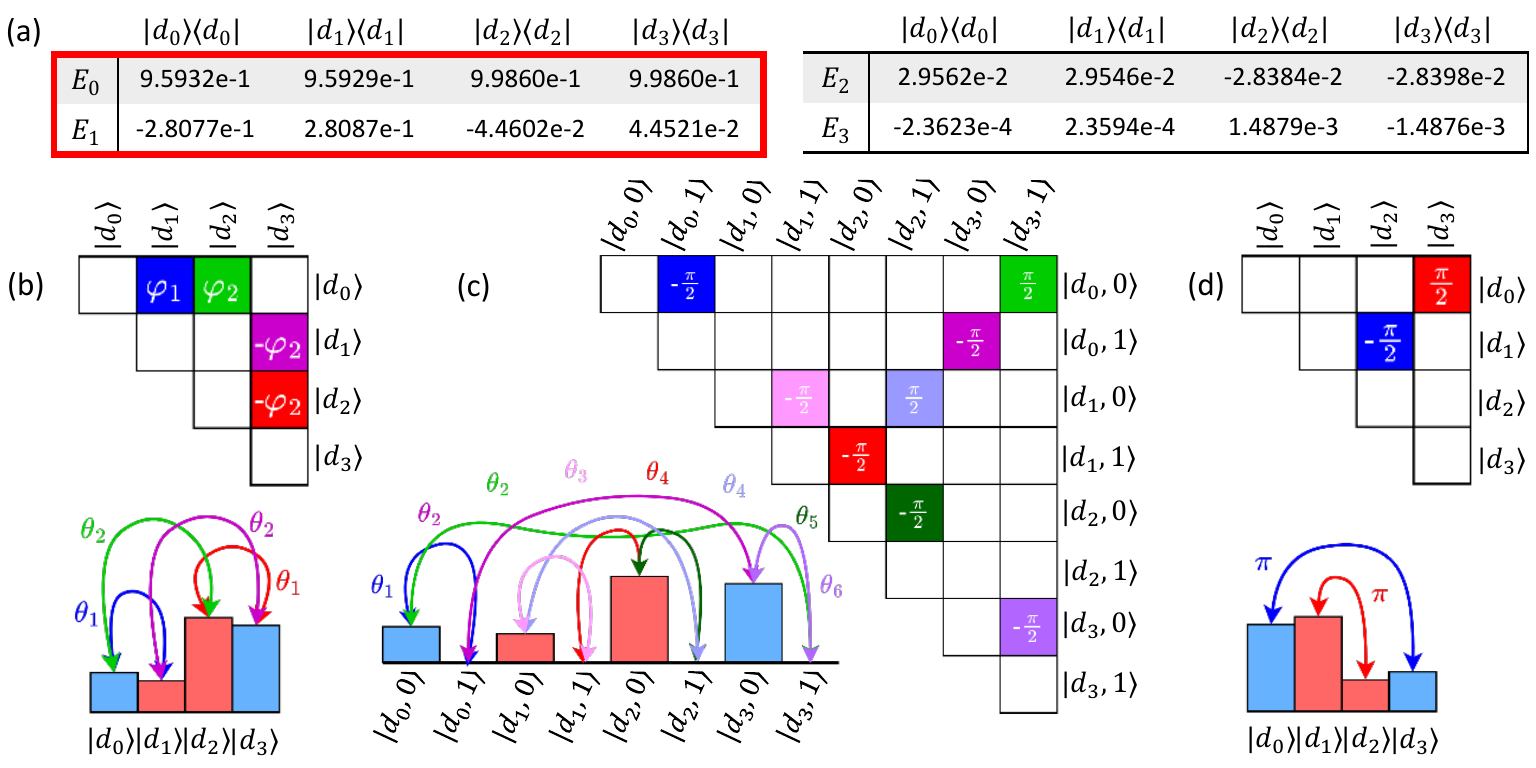}
    \caption{(a) Diagonal elements of the error operators $E_k$, $k=0,1,2,3$ for $d=4$, in the basis of the eigenvectors $\ket{d_k}$.     
    Scheme of single-qubit gates (b), stabilization (c), and recovery (d) for a $d = 4$ LQ, using the code words reported in Table II of the \ste{SM}.
    In each panel the histogram represents the populations of the eigenstates in blue (orange) for $\ell=0 (1)$. The upper triangular matrix on the right is the driving Hamiltonian in the rotating frame, with filled boxes indicating the set of $\theta_i$ resonant pulses between pairs of eigenstates (arrows in the histograms) which are simultaneously sent to the system. Different colors of the arrows and boxes indicate different pulse frequencies, while phases are indicated on top. For stabilization (c) the eigenstates of a two-level ancilla ($0,1$) are also reported.}
    \label{fig:scheme}
\end{figure*}

{\it Fault-Tolerant implementation --}
For a Fault-Tolerant implementation of the \textit{general protocol} discussed above the physical system used as a \textit{qudit} must fulfill two key properties.

First, we need the QEC code to propagate errors at a higher order than the elementary one. This requires  a well defined hierarchical structure in the Kraus representation of 
the error channel, since only at most $d/2$ of them can be handled by the code.

Second, each operation $U$ on the LQ  must be realized in an ET \cite{Ma2020} fashion, i.e. in such a way that the evolution between the error spaces (with different $\ell$) is exactly the same as that between the code spaces \cite{Ma2020}. 
This is achieved if $U$ is implemented in a {\it single step} by properly engineering the driving Hamiltonian $\tilde{H} = i \log U$. 
The latter corresponds to the coupling of the system with an external field, such as an oscillating electromagnetic field for a spin system.
The one-shot implementation of $U$ thus requires a specific connectivity between the qudit eigenstates, i.e. the possibility to {\it directly} induce transitions between them via resonant pulses. 
As detailed below, this in practice implies an all-to-all connectivity between the qudit eigenstates. 
This requirement can be met by encoding a qudit into a MSQ, thanks to the high degree of chemical tunability of the Hamiltonian in this class of systems \cite{Sessoli2019,Carretta2021,Lockyer2021,Lockyer2022,Rogers2022,Takui2012cnot,Rugg2019,Olshansky2020,Jacobberger2022}.

In particular, we apply the general formalism described above to a illustrative 7-spin cluster, introduced in Refs. \cite{Chiesa2022,Chizzini2022}.
The system Hamiltonian is the following:
\begin{equation}
H =  \sum_{i>j} J_{i,j} {\bf s}_i \cdot {\bf s}_j  + \sum_{i>j} D_{i,j} ({\bf s}_i \cross {\bf s}_j)_z + \mu_B B \sum_i g_i s_i^z,
    \label{eq:Ham}
\end{equation}
where $s_i^\alpha$ are spin operators on sites $i=1,...,7$, arranged to form a pair of corner-sharing tetrahedra as in the existing Ni$_7$ molecule \cite{Ni7}. 
In this molecule, the central spin is $s_7=3/2$, while all the others are $s_i=1/2$. 
In Hamiltonian \eqref{eq:Ham}, the first term is the dominant Heisenberg exchange interaction between neighboring spin pairs with anti-ferromagnetic couplings $J_{i,j}$, the second is the (axial) anti-symmetric exchange coupling and the third is the Zeeman interaction with an external magnetic field $B$ along $z$.
By proper choice of the hierarchy between the different parameters, this Hamiltonian gives rise to an energy spectrum consisting of eight low-energy total spin $1/2$ doublets well separated from the first excited $S=3/2$ (see \ste{SM}).
Moreover, the anti-symmetric exchange couples to first order all the different low-energy multiplets. {\it Hence, this molecule provides up to 16 low-energy eigenstates whose mutual transitions can be individually and directly addressed.} This feature is specifically exploited in the FT implementation of our protocol. It can be obtained also using simpler MSQs, such as single-spin molecules in presence of transverse magnetic fields and/or high-rank zero-field splitting anisotropy.

In MSQs (as in trapped ions \cite{RevTrappedIons,Ringbauer2022}, atoms \cite{Kuhr2005} and several other spin architectures \cite{Morello2020,NatureRevMorello}) 
pure dephasing (with decay time $T_2$) is the leading error  at low temperature, while spin relaxation is orders of magnitude slower \cite{Freedman_Cr,Bader2016,SIMqubit}.
Therefore in the following we will focus on it. 
Pure dephasing in MSQs arises from the coupling between the system spins and the surrounding nuclear spins \cite{BaderNatComm14,npjQI}. In the secular and Born-Markov approximations, one can derive a Lindblad equation for the evolution of the density matrix \cite{Chiesa2022}, resulting in a decay only of off-diagonal terms of $\rho$ [$\rho_{ij}(t) = \rho_{ij}(0) e^{-\gamma_{ij}t}$] with rates $\gamma_{ij} = \sum C_{kk'}^{zz} [ \bra{d_i} s_k^z \ket{d_i} \bra{d_i} s_{k'}^z \ket{d_i} + \bra{d_j} s_k^z \ket{d_j} \bra{d_j} s_{k'}^z \ket{d_j} -2 \bra{d_i} s_k^z \ket{d_i} \bra{d_j} s_{k'}^z \ket{d_j} ]$. These are essentially related to the structure of the eigenstates $\ket{d_i}, \ket{d_j}$, apart from the coefficients $C_{kk'}^{zz}$ which 
depends on the molecular structure (see \ste{SM}). 
From the solution of the Lindblad equation we extract the corresponding {\it diagonal} Kraus operators $E_k$ at a fixed time by state tomography \cite{Nielsen}. 
Since they display a clear hierarchical structure, we can focus on the $d/2$ most relevant ones and solve the KLc \eqref{eq:KL}. 
For instance, the $E_k$ operators for $d=4$ are reported in the top panel of Fig. \ref{fig:scheme}, in order of decreasing norm. $E_0$ and $E_1$ (red box) are clearly leading and we can fulfill the KLc for them. The magnitude of the components of the resulting code words in the eigenbasis is shown by the bars in Fig. \ref{fig:scheme}. 
We have tested with different model Hamiltonians and spatial distributions of nuclear spins that the hierarchical structure of the errors is rather general. 
The other $E_k$ operators and the corresponding code words are reported in the \ste{SM}, for different number qudit eigenstates ($d$ from 4 to 12) used to define the LQ. Note that by changing $d$ we are actually changing the code and hence a threshold analysis for each case is needed (see below). 

Based on this molecular spin hardware, we now discuss the ET implementation of logical gates.
Let us start from a logical planar rotation $R_L^P(\theta,\phi)= \exp \left[ -i \left( \cos \phi \; Y_L - \sin \phi \; X_L \right) \theta/2 \right]$, as sketched in Fig. \ref{fig:scheme}-(a). 
Its ET realization requires resonant pulses between each eigenstate in the logical $\ell=0$ subspace [blue bars in Fig. \ref{fig:scheme}-(b)] and each one in the logical $\ell=1$ subspace (orange bars). 
Different pulses of length $\theta_j$ are indicated by double arrows in the histogram and their phase $\varphi_j$ is reported in the connectivity matrix $\tilde{H}$ nearby. 
As detailed in the \ste{SM}, one can show (with a generalized rotating frame formalism \cite{Leuenberger2003}) that the set of simultaneous pulses $\{\theta_j,\varphi_j\}$ implements the unitary $\exp\small[-i \tilde{H}\small]$, which corresponds to  $R_L^P(\theta,\phi)$.
Here we assume all the energy gaps to be spectroscopically distinguishable.
Remarkably, since $\theta,\phi \in \mathbb{R}$, the resulting set of gates $R_L^P(\theta,\phi)$ is a universal set for one-body logical operations.
Along the same lines one can implement in a single step also a two-body ET $C-\varphi$ logical gate (see below).

We now examine the stabilization process, reported in Fig. \ref{fig:scheme}-(c). 
In this case, the eigenstates of both the LQ and of a $d/2$ levels ancilla are shown in the product basis $\ket{d_j,k}$. 
We recall that the ancilla is initialized in its ground state $\ket{0}$. 
$CU$ stabilization requires to excite the ancilla iff the LQ is in $k=1$. 
This translates into the set of simultaneous pulses indicated in Fig. \ref{fig:scheme}-(b) between each qudit eigenstate  $\ket{d_i}$ with the ancilla in $\ket{0}$ and each qudit eigenstate $\ket{d_j}$ with the ancilla in $\ket{1}$. 
Following the argument above, one can derive the connectivity matrix with pulse lengths and phases by taking the logarithm of the $CU$ gate (see \ste{SM}). Remarkably, the ancilla does not need to be encoded. Indeed, it is excited to the eigenstate $\ket{k}$ with the same probability $p_k$ of having an error $E_k$.
Hence, it suffers an error $E_{k^\prime}$ with the conditional probability  $p_{k} \, p_{k^\prime}$ which therefore results to be of a higher order than the error on the \textit{qudit}. 
Fig. \ref{fig:scheme}-(d) also shows the recovery step from $k=1$ to $k=0$ (for the specific case $d=4$), which is obtained by a pair of pulses within the two disjoint $\ell=0,1$ subspaces \footnote{This is due to the specific form of the error words $\{ \ket{\ell, k}\}$, in which differ the two $\ell=0,1$ subspaces are disjoint for all $k$.}. \\
Summarizing, we have presented a MSQ characterized by a clear  hierarchical structure in the error operators and providing the capability to implement in a single step any QC operation of the embedded QEC protocol. Hence, the proposed implementation  of the code is Fault-Tolerant.

\begin{figure}[t!]
    \centering
    \includegraphics[width=0.475\textwidth]{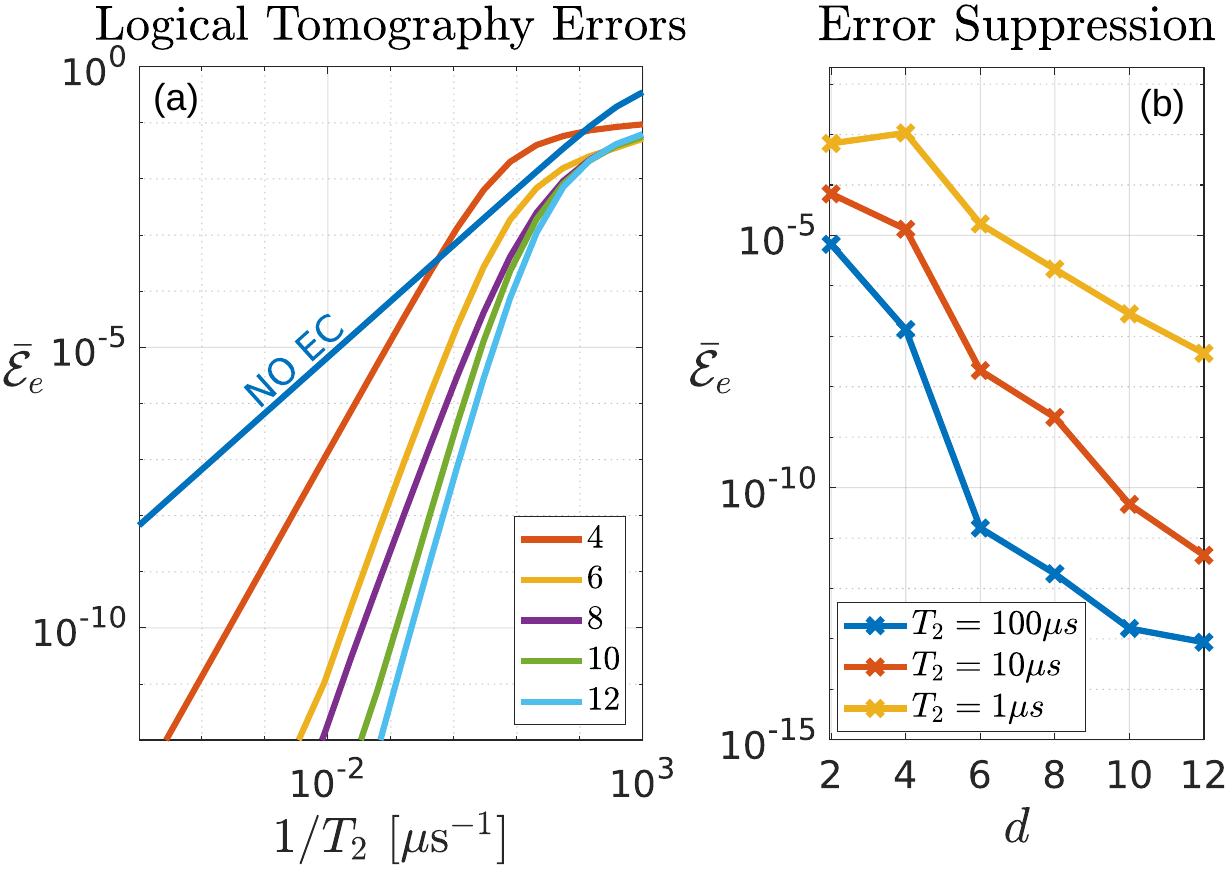}
    \caption{Simulation of the Fault-Tolerant implementation of single qubit logical gates on a MNM \cite{Chiesa2022}. (a) Error $\mathcal{E}_e = 1 - \mathcal{F}_e^2$ as a function of $1/T_2$, where $\mathcal{F}_e$ is the entanglement fidelity \cite{Knill2001}, averaged over different planar rotations $R_L^p(\theta,\varphi)$ for $(\theta, \phi) = (\frac{\pi}{4}, \pi), (\frac{\pi}{2}, \pi), (\frac{\pi}{2}, -\frac{\pi}{2}), (\frac{\pi}{2}, -\frac{\pi}{4}), (\frac{\pi}{2}, -\frac{\pi}{8})$. (b) Performance of the code as function of the number of levels for different values of $T_2$. The same driving field amplitude is assumed for manipulating the LQ and the uncorrrected qubit, represented by a spin 1/2 with $g=2$.}
    \label{fig:sim1q}
\end{figure}

{\it Threshold Analysis --}
We now show the effectiveness of our scheme for the correction of pure dephasing errors (with characteristic time $T_2$) in MSQs.
This is done by numerically integrating the Lindblad equation  for the system density matrix, subject both to pure dephasing and to the sequence of pulses required to implement the desired operations (see \ste{SM}). These include a universal set of logical gates, stabilization ($CU$), recovery. We also consider possible measurement errors (see below).
Differently from standard threshold analysis \cite{Mariantoni2012}, we are not assuming discrete errors occurring with a given probability but a continuous dephasing realistically describing the  system and acting during all the procedures, with elementary error rate $1/T_2$. 
\footnote{This allows us to compare the performance of a spin 1/2 qubit (without EC) and of the LQ even though the duration of the elementary transitions on them  is different (and hence the related error probability). Indeed, the elementary error probability for a two-level system is $p=(1-\exp[-\tau/T_2])/2$ \cite{Nielsen}. Since the duration $\tau$ of the transition  is much smaller for a spin 1/2, it would result in a much smaller $p$ compared to the LQ, thus making the comparison unfair.}).

We start by considering generic $R_L^P(\theta,\phi)$ rotations, followed by stabilization and correction and we report [Fig. \ref{fig:sim1q}-(a)] the final {\it logical} error $\bar{\mathcal{E}}_e$ as a function of the elementary error rate $1/T_2$. 
Here $\mathcal{E}_e = 1 - \mathcal{F}_e^2$, where $\mathcal{F}_e$ is the entanglement fidelity \cite{Knill2001} of the procedure \footnote{$\mathcal{F}_e$ is the average fidelity on the initial states $\ket{0_L}, \ket{1_L},$ $\frac{1}{\sqrt{2}}(\ket{0_L} + \ket{1_L}), \frac{1}{\sqrt{2}}(\ket{0_L} - \ket{1_L}), \frac{1}{\sqrt{2}}(\ket{0_L} + i\ket{1_L}), \frac{1}{\sqrt{2}}(\ket{0_L} - i\ket{1_L})$}, and we average on a universal set of $(\theta,\phi)$ values. 
The {\it logical} error is obtained by full state tomography of the logical state and corresponds to the failure probability of the QEC protocol, i.e. to  errors which cannot be corrected by the code.

We immediately note that the slope of the curves in Fig. \ref{fig:sim1q}-(a) in log-log scale for the LQ (number of levels $\ge 4$) is larger than for the uncorrected case (blue line). 
This means that $\bar{\mathcal{E}}_e$ is propagated through the executed circuit to a higher order than the elementary error, thus remarking the Fault Tolerance of all the procedures involved.
Notably, for reasonable gate duration ($90$ ns for the LQ, see  \cite{Atzori_JACS,Atzori2016,VCp2Cl2} and \ste{SM}) the LQ with $d=4$ beats the uncorrected two-level system for $T_2 \gtrsim 2 \;\mu$s, a perfectly achievable values in MSQs \cite{BaderNatComm14,Atzori_JACS,SIMqubit}. 
This corresponds to an error probability for the elementary transition between a pair of qudit levels of  $p \gtrsim \tau/2T_2\approx 2.2\%$. 
By increasing $d$ (i.e. including more qudit eigenstates in the encoding) the codes beat the spin 1/2 {\it independently of $T_2$}. 
Indeed, errors are reduced and propagated to higher orders, as evidenced by the increase of the slope of the curves in Fig. \ref{fig:sim1q}-(a).

Remarkably, we find an almost exponential suppression of the error with a linear increase of $d$, as reported in Fig. \ref{fig:sim1q}-(b), independently of $T_2$. 
This makes the proposed FTQC scheme much more efficient than standard qubit concatenation techniques, where an exponential error suppression is obtained only with an exponential growth of the Hilbert space \cite{Nielsen}.
To quantify this gain, we note that the final error suppression is strikingly large when compared to standard multi-qubit codes: here we achieve a logical error of $10^{-12}$ by employing $12 \times 6$ levels (LQ + ancilla) and $T_2=10 \; \mu$s, which should be compared to more than $5000$ qubits needed by Floquet codes with the same elementary error $p\approx \tau/2T_2 = 0.45\%$ \cite{Paetznick2023}.  
This remarkable performance, obtained exploiting \textit{qudits} with all-to-all connectivity, is here shown by focusing on the dominant family of errors of molecular \textit{qudits}. 

\begin{figure}[t!]
    \centering
    \includegraphics[width=0.47\textwidth]{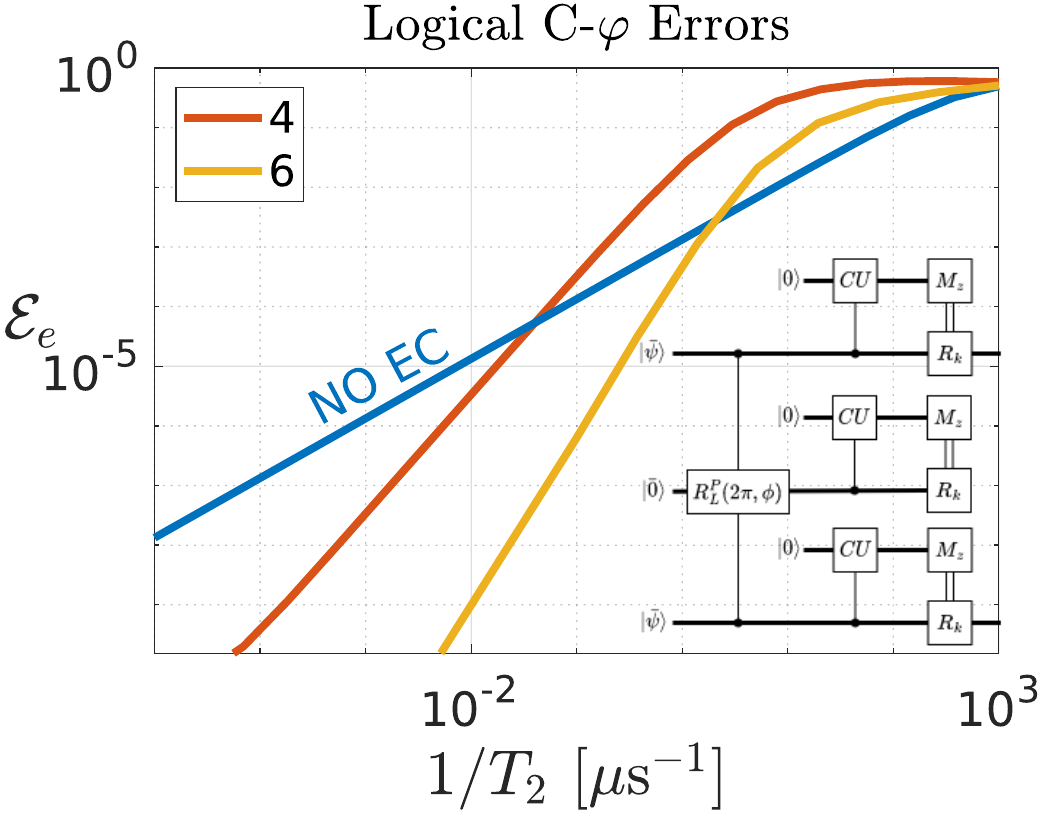}
    \caption{Numerical simulation of the FT implementation of the two-qubit logical $C$-$\varphi$. The logical error is shown as a function of $1/T_2$  at the end of a ‘‘logical gate - EC" cycle. Simulations are computationally demanding and hence we limit to $d=4, 6$.
    Inset: simulated circuit, with external lines corresponding to LQs in a  generic logical state $\ket{\bar{\psi}}$, coupled by an error-corrected switch initialized in  $\ket{0_L}$ (middle line). Each unit is coupled to a $d/2-$levels ancilla for stabilization.}
    \label{fig:sim2q}
\end{figure}

Leakage errors could arise due to the finite duration of the pulses, but can be largely reduced by pulse-shaping techniques \cite{Castro2022} and/or by using longer pulses.
Indeed, increasing the length of the pulses simply shifts the curves in Fig. \ref{fig:sim1q} to the left without changing their slope. 
This will slightly increase the threshold value of $T_2$, without compromising Fault-Tolerance.

Moreover, the {\it depth of our circuit does not increase with $d$}, since all operations are implemented in parallel, again in contrast with standard codes.
Remarkably, here a single measurement is sufficient for syndrome extraction and the corresponding error \cite{vqe1} can be suppressed by repeated measures as shown by simulations in the \ste{SM}.

We now consider a two-qubit logical gate, such as the controlled-phase $C-\varphi$. In the MSQ architecture, this can be realized by considering a molecule consisting of three interacting LQs, where the middle one acts as a switch of the effective coupling between the other two \cite{modules,AIPadv}, $Q_1$ and $Q_2$ (inset of Fig. \ref{fig:sim2q}). 
In presence of an interaction between the three units, excitations of the switch are dependent on the state of both $Q_1$ and $Q_2$. 
This allows us to induce a $2\pi$ (semi-)resonant excitation of the switch (initialized in $\ket{0_L}$) only for a specific 2-qubit {\it logical} state, such as $\ket{1_L 1_L}$.
As a result, a phase is added only to the $\ket{1_L 1_L}$ component of the two-qubit logical state \cite{SciRepNi,modules,AIPadv,Blueprint_}, i.e. a $C-\varphi$ gate is implemented (see also \ste{SM}).\\
Since this involves a logical rotation $R_L^P (2\pi,0)$ of the switch, the latter must also be encoded and corrected.
Simulation of the two-qubit gate followed by EC on all the three units is shown in Fig. \ref{fig:sim2q} for $d=4, 6$.
The logical error is suppressed for $T_2 \gtrsim 25 \; \mu$s ($4$-levels encoding) and for $T_2 \gtrsim 500$ ns ($6$-levels encoding), yielding $p \gtrsim \tau/2T_2 \approx 8\%$. We stress that these numbers are within experimental capabilities \cite{BaderNatComm14,Zadrozny2015,Atzori_JACS}.

{\it Conclusions --}
We have shown a Fault-Tolerant protocol for quantum computation with qudit embedded stabilizers codes in a molecular spin architecture \cite{Blueprint_}. 
This embedded protocol prevents the blowing up in the number of physical units, while yielding a large error suppression already with a small number of \textit{qudit} levels, thus making the proposed route very appealing for an actual implementation. 
Moreover, we have pinpointed the crucial requirements for a physical system to achieve FTQC, thus potentially extending the strategies proposed here to other qudit platforms showing a well defined hierarchy of errors and proper connectivity between its eigenstates.

\acknowledgements
$^\dagger$These Authors contributed equally to the work.

We warmly thank P. Santini for useful and stimulating discussions.

This work received financial support from European Union – NextGenerationEU, PNRR MUR project PE0000023-NQSTI, from the European Union’s Horizon 2020 program under Grant Agreement No. 862893 (FET-OPEN project FATMOLS), from Fondazione Cariparma and from 
Novo Nordisk foundation under grant
NNF21OC0070832 in the call “Exploratory Interdisciplinary Synergy Programme 2021”.
Project funded by the European Union – NextGenerationEU under the National Recovery and Resilience Plan (NRRP), Mission 4 Component 1 Investment 3.4 and 4.1. Decree by the Italian Ministry n. 351/2022 CUP D92B22000530005.

%

\end{document}